 \documentclass[showpacs,preprint,aps]{revtex4}
\usepackage{epsfig}
\begin{document}

\title{
    Low energy shape oscillations of negative parity in the main and
   shape-isomeric minima in actinides
         }

\author{M.~Kowal}\email{mickowal@fuw.edu.pl}
\author{J.~Skalski} \affiliation{ Soltan Institute for Nuclear Studies,
Ho\.za 69, PL-00-681 Warsaw, Poland \\}


\begin{abstract}

  We study low energy shape oscillations of negative parity in the first and
  second (isomeric) minima in actinides. As a main tool we use the
 phenomenological Woods-Saxon potential with a variety of shape deformations.
  This allows to include a mixing of various multipolarities when considering
  oscillations with a fixed $K$ quantum number. The phonon energies are
  determined either from the collective Hamiltonian with
  the microscopic-macrocopic energy and cranking mass parameters, or from
  its simplified version with the constant mass parameters.
   The results for $K^{\pi}=0^-$,$1^-$ in the first minima are in a reasonable
  agreement with experimental data, including predicted E1 transitions;
  the $K^{\pi}=2^-$ energies are systematically overestimated.
  In the second minimum, as compared to the data for $^{240}$Pu and $^{236}$U,
   our calculated $K=$1,2 energies are overestimated while the $K=0$ energies
   are three or more times too large.
  This signals either a non-collective character of the experimentally
  assigned $K=0$ states or a serious flaw of the model in the second minimum.
  More data on the $K=0$, $I^{\pi}=1^-$ collective states in the second minima
  of other nuclei are necessary to resolve this issue.

\end{abstract}


\maketitle


\section{Introduction}

  Recently, a considerable amount of data has been gathered on the nuclear
 states in the second well in the actinide region, especially
 in $^{240}$Pu, see \cite{Thirolf} and references cited therein.
 Various rotational bands, most of them of negative parity,
  have been identified in this nucleus with moments of inertia
 characteristic of superdeformation (SD).
 Such spectroscopic data provide a much wanted test for nuclear models that
 were originally fitted in a region of deformations around the ground state
  (g.s.) minima:
  How much of their predictive power is left beyond that region?
  Negative-parity shape oscillations are natural candidates for the observed
  low energy band-heads at the SD.
   Here we study them within the much used microscopic-macroscopic
  model based on the deformed Woods-Saxon potential.
  The found properties of excitations in the shape-isomeric
  minima provide both a prediction and a test of the model.
  We check the used method by first applying it to the negative-parity
  excitations
  in the first minima in actinides, on which data are more abundant.

Discovery of fast fissioning states in actinides by Polikanov et
al.
 \cite{Pol} was interpreted soon afterwards by Flerov and Druin \cite{FD66}
 and Strutinsky \cite{St67} as evidence for very deformed secondary minima
 in these nuclei.
 Later experiments provided support for this guess, with measurements of
 rotational band in $^{240}$Pu \cite{Specht} and then, of the quadrupole moment
 in the second well in $^{239}$Pu \cite{Rev}.
 However, very short lifetimes, in the range from 10 ps to 10 ms, and
 experimental difficulties precluded, and in fact preclude to this very day,
 gainining sufficient knowledge on the nuclear structure in the second
  minimum.
  Only four values of the quadrupole moment in the second well were measured
  via nuclear methods of the total of 34
 shape-isomeric states. Three additional data for Am isotopes come from
  the optical isotope shift and hyperfine structure measurements of the
 difference between the mean square radii in the ground and isomeric states
  \cite{Backe80}.
Moreover, difficulties in the experimental access to fission
isomers make
 the results very dependable on rather extended argumentation chains
 involving assumptions instead of established experimental facts.
 Although this does not necessarily invalidates the claimed results,
 quite a number of legitimate reservations may be formulated which diminish
  their firmness, see \cite{Makar}.


  We determine properties of low-lying negative-parity shape vibrations in
  even-even actinides using the collective model,
 i.e. the Schr\"odinger equation with deformation
  parameters as coordinates. As the ingredients we take the
  microscopic-macroscopic energy and cranking mass parameters.
  In order to describe nonaxial modes we include
  nonaxial deformations of the nuclear shape corresponding to $K=1,2,3$.
  The use of the microscopic-macroscopic energy
   is consistent with the working assumption that it correctly
  predicts the shape dependence of energy for "cold" nuclear configurations.
  On the other hand, using adiabatic mass parameters is only an
  approximation. It should work reasonably well
   for phonon energies smaller than $2 \Delta$,
   with $\Delta$ the neutron or proton pairing gap,
   The experimental energies in the second well and many in the first well
  fulfil this condition \cite{Thirolf}.

   One might expect that negative-parity shape vibrations are
   mostly octupole. However, there are at least two reasons to
  consider admixtures of higher order odd-multipolarity modes, i.e.
   $\lambda=5$ and 7: 1) The equilibrium shapes are spheroidal, hence the
  multipole components of different $\lambda$ are not the normal modes (not even
  being orthogonal) and may couple with each other, 2) Deformation parameters
  $\beta_{\lambda \mu}$ of the
  microscopic-macroscopic model define the s.p. potential, not the density.
  Therefore, the octupole part of the latter may be produced by various
  odd-$\lambda$ deformations of the potential. As a consequence,
   one should account for a possible mode coupling when looking for low lying
  excitations of negative parity.
  We are not aware of any other study of shape oscillations explicitely
  including odd-rank multipoles of the order higher than three.

 One can observe that an alternative approach to the study of octupole
 vibrations,
 the schematic Random Phase Approximation \cite{NerV1,RPAJ,RPAR}, while free
 from the adiabaticity assumption that we use for mass parameters, has its own
  problems, including the necessity of fixing the coupling constants.
  In order to study the coupling of various multipolarities, one would have
  many constants to fix.

\section{Method}

  We use a sigle-particle (s.p.) Hamiltonian with the deformed Woods-Saxon
 potential defined in terms of the nuclear surface, according to the scheme
 exposed in \cite{WS}. However, the potential used in the present work admits
  more general nuclear shapes:
  The only restriction imposed on them is that they have one symmetry plane
  $y$-$z$.
  This leaves one conserved signature quantum number, $s_x=\pm i$, being the
  eigenvalue of the signature operator
  ${\cal S}_x={\cal P}{\cal R}_x^{-1}$, with ${\cal P}$ the intrinsic parity
 and ${\cal R}_x$ the rotation by $\pi$ about the intrinsic $x$ axis.
  The degeneracy of pairs of states with $s_x=\pm i$ (the Kramers degeneracy)
   reduces by half the dimension of the s.p. Hamiltonian matrix.

  Nuclear shapes compatible with the assumed symmetry are
  defined by the following equation of the nuclear surface
\begin{eqnarray}
  \label{shape}
    R(\theta,\varphi) &=& c(\{\beta\})R_0 (1+ \sum_{\lambda>1}\beta_{\lambda 0} Y_{\lambda 0}(\theta,\varphi) \nonumber\\
   &+& \sum_{\lambda>1, \mu>0, even}\beta_{\lambda \mu} Y_{\lambda \mu c} (\theta,\varphi) \nonumber\\
   &+& \sum_{\lambda>1, \mu>0, odd} \beta_{\lambda \mu} Y_{\lambda \mu s} (\theta,\varphi))  ,
 \end{eqnarray}
 where $c(\{\beta\})$ is the volume-fixing factor. The real-valued spherical
 harmonics, $Y_{\lambda \mu c}$ with even $\mu>0$
   and $Y_{\lambda \mu s}$ with odd $\mu>0$, are defined in terms of the
  usual ones as: $Y_{\lambda \mu c}=(Y_{\lambda \mu}+Y_{\lambda -\mu})/
 \sqrt{2}$ and $Y_{\lambda \mu s}=-i(Y_{\lambda \mu}+Y_{\lambda -\mu})/
 \sqrt{2}$.
  In other words, the dependence of the shape on the azimuthal angle $\varphi$
  enters through functions $\cos(\mu \varphi)$ with $\mu$ even and
  $\sin(\mu \varphi)$ with $\mu$ odd.

  For the macroscopic part we used the Yukawa plus exponential model
  \cite{KN}.
   The parameters of both the macroscopic part and the s.p. potential
   used in the present work, as well as the way
  the shell- and pairing corrections are calculated, are the same as in
  a number of previous studies, e.g. in \cite{WSpar}.


   \subsection{Oscillations around I and II minima}

  The second minima in actinides found in this and previous calculations,
  see e.g.  \cite{Stefan,Delar}, correspond to the axially- and
 reflection-symmetric shapes.
  The same holds for the g.s. minima, except for some light thorium,
 uranium and plutonium isotopes with $N \leq 138$ that have octupole
 equilibrium deformations and are {\it not included} in the present study.
 So we are left with nearly parabolic, not too shallow minima.
  Small oscillations around them with different intrinsic $K$ numbers or
  parities are nearly independent. Indeed, the
  amplitude of vibrations cannot be too large, while exactly at the minimum
  the modes are uncoupled, except for the Coriolis coupling which may be then
  considered as a perturbation.

  The treatment of negative-parity shape oscillations in the collective
  model is based on the collective Hamiltonian
 \begin{equation}
  \label{Ham}
 {\hat H} =   -\frac{\hbar^2}{2}\frac{1}{\sqrt{\det B}} \sum_{i,j=1}^3
  \frac{\partial}{\partial \beta_i}
  \left(\sqrt{\det B} (B^{-1})_{i j}\frac{\partial}{\partial \beta_j}\right)
   + V(\beta_k)
  \end{equation}
  diagonalized  within the space of collective wave functions with the
  scalar product $\langle \psi_1\mid\psi_2\rangle=\int d^3\beta_k
  \sqrt{\det B} \psi_1^* \psi_2$. Here $V(\beta_k)$ is the
  microscopic-macroscopic energy and
  $B_{i j}(\beta_k)$ is the mass tensor with indices corresponding to
   deformations $\beta_{\lambda K}$, $\lambda = 3,5,7$. In the present work
   we use cranking mass parameters.
  Oscillations for each $K=0,1,2,3$ are considered
  separately. A similar approach was used in \cite{collPom} to study
  the $K=0$ octupole state in $^{222}$Ra. In contrast to that
  work, where both quadrupole and octupole coordinate were used, we restrict
   collective variables to reflection-asymmetric deformations.

  Since our study is confined to nuclei with sufficiently deep, nearly
 parabolic minima, we can use the approximation of small oscillation
 amplitudes. This formalism follows from the one above
  if we replace $V$ by the quadratic form
  $(1/2)\sum_{i j} C_{i j} \beta_i \beta_j$,
   approximately valid around the potential minimum, with the stiffness
  coefficients $C_{i j}$, and fix the mass parameters at the values
  calculated at this minimum. Then the Hamiltonian becomes
 \begin{equation}
 \label{Hams}
 {\hat H} =   -\frac{\hbar^2}{2} \sum_{i,j=1}^3
   (B^{-1})_{i j}\frac{\partial^2}{\partial \beta_i \partial \beta_j}
   + \frac{1}{2} \sum_{i,j=1}^3 C_{i j} \beta_i \beta_j  .
  \end{equation}
 Within this 3D harmonic oscillator model, the study of the $\lambda=3,5,7$
 coupling is straightforward. The eigenmodes $\xi_k$ are given by a
 transformation of coordinates
  \begin{equation}
  \label{coor}
   \beta_i =  \sum_{k=1}^3 \left(\sum_{j=1}^3
 \sqrt{\frac{\hbar}{B_{D j}\Omega_{D k}}}
  S_{1 i j} S_{2 j k}\right) \xi_k ,
  \end{equation}
  where the orthogonal matrices $S_1$ and $S_2$ diagonalize the mass tensor,
  $S_1^T B S_1 = B_D$, and the frequency matrix $\Omega$, $S_2^T \Omega S_2 =
  \Omega_D$, with $\Omega^2 = B_D^{-1/2} S_1^T C S_1 B_D^{-1/2}$.
   The square of the oscillation frequency $\hbar \omega_K$ corresponding
   to the oscillation with a given $K=$0,1,2,3 is the smallest solution to
   the cubic equation
   $\det(C_{i,j}-(\hbar\omega_K)^2 B_{i j})=0$, with $C_{i j}$ and $B_{i j}$
   the stiffness and mass 3$\times$3 matrices, $i,j=\beta_{\lambda K}$ and
   $\lambda=3,5,7$.

  The calculations of the coupled oscillations with the Hamiltonian
  (\ref{Ham}) were made only for selected cases, as they require
    a time-consuming calculation of mass parameters, especially for
   $K\neq 0$. Some technical aspects of these calculations,
   as well as the evaluation of the cranking mass parameters
   for the rich deformation set Eq. (\ref{shape}),
   are shortly described in the Appendix A.
  As discussed in Sect. III below, the results obtained with deformation-
   dependent mass parameters are not very different from those of
  the much simpler version with constant mass parameters (\ref{Hams}).

  \subsection{Electric dipole transitions}

  Reduced probabilities of electromagnetic (EM) transitions between the
  rotational band built on the one-phonon state and the g.s. band can be
 calculated assuming the fixed structure of both the phonon and the collective
  rotor \cite{BM}. For an operator ${\cal M}$ of the multipolarity $\lambda$
 one has
  \begin{eqnarray}
  \label{Btran}
  B(\lambda;K_1=0,I_1\rightarrow K_2,I_2)& = &
 (2-\delta_{K_2 0}) \langle I_1 0 \lambda K_2\mid I_2 K_2\rangle^2
  \mid\langle K_2\mid{\cal M}(\lambda,K_2)\mid K_1=0\rangle\mid^2
 \end{eqnarray}
   with ${\cal M}(\lambda,K_2)$ the intrinsic spherical component.
 For the negative-parity shape vibrations, the most prominent are dipole
  transitions, if not hindered by special reasons, as the octupole transitions
 are usually much weaker. It is especially true for shape isomers where
  only E0 and E1 transitions are observed \cite{Gassmann}. We have ${\cal M}(E1,0)=[3/(4\pi)]^{1/2}
 {\hat D}_z$ and ${\cal M}(E1,1)=-[3/(4\pi)]^{1/2} ({\hat D}_x+ i {\hat D}_y)
/\sqrt{2}$, where
 the dipole moment ${\hat {\bf D}}= e(N\sum_p {\bf r}_p-Z\sum_n {\bf r}_n)/A$.
 Since the $K=1$ phonon is the equal-weight combination of the equal-energy
 phonons in directions $x$ and $y$, the intrinsic matrix element in Eq.
 (\ref{Btran}) equals $-[3/(4\pi)]^{1/2} D^t_y$, where $D^t_y=\langle
 Y_{\lambda 1 s} \mid {\hat D}_y\mid g.s.\rangle$, with symbolically denoted
  one-phonon state induced by the deformations
 $Y_{\lambda 1 s}$, $\lambda=3$,5,7, Eq.(\ref{shape}).

    Transition matrix elements in the intrinsic frame between the g.s.
   and the lowest excited state of negative parity $\mid \pi -\rangle$
   could be calculated by integrating the transition density
   $\rho_{tr}(\beta_k)= \{\sqrt{\det B} \psi^*_{\pi-}
   \psi_{gs}\} (\beta_k)$ with the proper operator, represented in the
   collective space. Taking the dipole moment as an example,
   we have
  \begin{equation}
  \label{Dt}
  D^t=\langle \pi -\mid {\hat D}\mid 0\rangle = \int_{-\infty}^{\infty}
  d^3\beta_k \rho_{tr}(\beta_k) D(\beta_k) ,
   \end{equation}
  where $D(\beta_k)$ is the expectation value
  of ${\hat D}$ in the mean field state with the deformations ${\beta_k}$
  \cite{collPom}.

   A considerably simpler approximation consists
   in calculating the diagonal matrix element of the transition operator in
   the mean-field state with the deformations $\beta^{tr}_k$ fixed as the
   most probable by the above transition density:
   \begin{equation}
 \beta^{tr}_k = \int_{-\infty}^{\infty} d^3\beta_j \rho_{tr}(\beta_j) \beta_k .
   \end{equation}
   The dipole and octupole moment operators are to the leading order linear in
  the odd-multipole deformation parameters $\beta_{\lambda K}$, so indeed, the
   integration in Eq. (\ref{Dt}) replaces $\beta_{\lambda K}$ in
  $D(\beta_{3 K},\beta_{5 K},\beta_{7 K},...)$ by $\beta^{tr}_{\lambda K}$.
  A nonlinearity of $D$ as a function of deformations may introduce
  an error in this approximation.
  Assuming that the lowest negative-parity mode
   corresponds to the normal coordinate $\xi_1$, the values
    of $\beta^{tr}_k$ for {\it harmonic} vibrations are actually given by
  Eq.(\ref{coor}) with $\xi^{tr}_1 = 1/\sqrt{2}$, $\xi_2=\xi_3=0$.

  This may be contrasted with the strong coupling limit with two octupole
  minima, at $\pm \beta^{eq}_{\lambda K}$, in which the transition
   matrix element $D^t$ is calculated as the expectation value at this
   deformation of equilibrium. One can notice that in this case,
   $\beta^{eq}_{\lambda K}$ is equal to $\beta_{\lambda K}^{\pi-}$, the
  expectation value of
  $\beta_{\lambda K}$ in the first excited state of negative parity,
  nearly degenerate with the g.s., $\beta_{\lambda K}^{\pi-}=
  2\int_0^{\infty}\sqrt{\det B}
 \mid \psi_{\pi-}\mid^2 \beta_{\lambda K} d^3\beta_k$.
  For the harmonic lowest-lying phonon one has
   the relation $\beta_{\lambda K}^{tr}=0.63
  \beta_{\lambda K}^{\pi-}$ which follows from Eq. (\ref{coor}) and
 the relation for the one-dimensional harmonic oscillator:
  $\xi^{tr}_1 = (\pi/8)^{1/2} \xi^{\pi-}_1 \approx 0.63 \xi^{\pi-}_1$,
  as $\xi^{\pi-}_1 = 2/\sqrt{\pi}$.

  Expectation value of the electric dipole moment in the state with
  deformations $\beta_{\lambda K}$, $K=$0,1, is calculated as a sum of the
  macroscopic and shell-corection parts, see e.g. \cite{Lean,ButNaz}. The
  macroscopic part, derived within the Droplet Model in \cite{DMS}, has to
  be calculated as in \cite{dip}, i.e. without assuming small
  $\beta_{\lambda \mu}$. For the completeness of the presentation, the
  relevant formulas and parameters are collected in the Appendix B.

 \subsection{Estimate of the Coriolis coupling effect}

  Within the rigid rotor-vibration coupling model, there are two modifications
 of the energy of stationary states with the total angular momentum $I$
 relative to the energy of one-phonon states: (i) a shift by
  $a(I(I+1)-K^2)$, with $a=\hbar^2/(2{\cal J})$, ${\cal J}$ the moment of
  inertia, and (ii) the effect of the Coriolis coupling
\begin{eqnarray}
 \label{Cor}
 \langle I M;K+1 \mid {\hat H}_{Cor}\mid I M;K\rangle  &=&
 -a \sqrt{1+\delta_{K 0}} \sqrt{(I-K)(I+K+1)}\langle K+1\mid {\hat J}_+\mid K\rangle ,
  \end{eqnarray}
   connecting states differing by $\Delta K = 1$. In the spherical limit,
  the matrix elements of ${\hat J}_+$ between substates $K$ and $K+1$ of a
  collective vibration of the multipolarity $\lambda$ are equal to
 $\sqrt{(\lambda-K)(\lambda+K+1)}$. The actual matrix elements for octupole
  phonons in deformed nuclei were found close to the spherical limit in
 \cite{NerV1}. Here we also include multipoles $\lambda=5$ and 7, so
   we estimate the effect of the Coriolis coupling using the spherical
  value with $\lambda$ consistent with the composition of the lowest-lying
  phonon, see sect. III.

 For a test case of $I=1$ states (the mixing of $K=0$ and $K=1$)
 in $^{240}$Pu, in the first minimmum, we take $\lambda=3$ and with
  $a=7.156$ keV \cite{Schmorak} find the coupling
  $H_{Cor}$=49.6 keV; in the second well, with $a=3.33$ keV \cite{Thirolf}, we obtain
  $H_{Cor}$=23.1 keV for $\lambda=3$ and 36.5 keV for $\lambda=5$.
  As the energy shifts of $I=1$ states due to the coupling
  (\ref{Cor}) are smaller then the coupling itself, especially when the
   difference in Coriolis-unperturbed energies is much larger than $H_{Cor}$,
   which is often the case, we do not include them in presented results.

\begingroup
\squeezetable

\section{Results and discussion}

 We begin with the description of energy minima obtained by a minimization
 in a multidimensional deformation space. A typical plot of energy vs.
 deformation $\beta_{20}$ after minimization with respect to other axial
 deformations is shown in Fig. \ref{fig:Pu240barier} for $^{240}$Pu.
 It is normalized by setting the macroscopic energy to zero
 at the spherical shape. The experimental excitation energy of the
 superdeformed state, determined by the statistical analysis, equals
  $E^{exp}_{II}=2.25$ MeV \cite{Thirolf}, whereas our calculation gives
  $E^{th}_{II}=2.0$ MeV. The calculated first and second barrier,
  $E^{th}_{A}=6.4$ MeV and $E^{th}_{B} =5.0$ MeV, are in a
   satisfactory agreement with the experimental values $E^{exp}_{A} =5.8$ MeV
 and $E^{exp}_{B} =5.45$.
  As mentioned before, mass-asymmetric and nonaxial deformations
  do not change the first and second minima.

  One can also see in Fig. 1 the calculated
  third minimum. A detailed discussion of the third minima goes beyond
 the scope of the present paper. Here, we can mention that nonaxial
  deformations, not included in Fig. 1, modify the energy surface around
  them.

\begin{figure}[h]
\vspace{-5mm}
\begin{minipage}[l]{0mm}
\centerline{\includegraphics[scale=0.90]{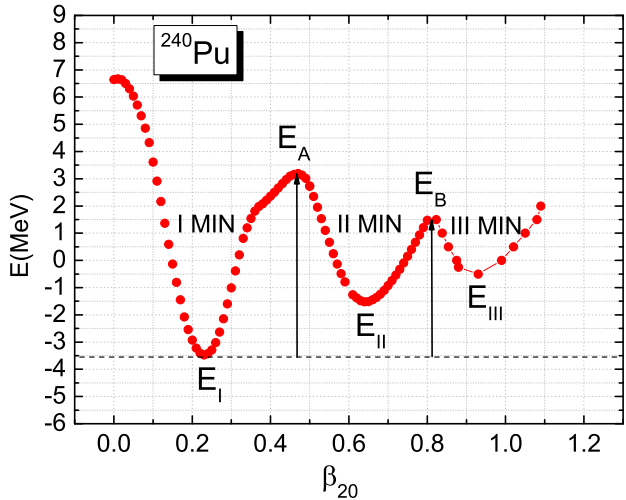}}
\end{minipage}
\caption{{\protect Calculated fission barrier for $^{240}$ Pu as a
function of $\beta_{20}$.
  }}
  \label{fig:Pu240barier}
\end{figure}

 Energies of shape vibrations depend on the values of the
  mass and stiffness parameters in the vicinity of the equilibrium
 deformation. The stiffness coefficients were determined by
   fitting energy by a quadratic form in deformations. A quality of this
   approximation in the first and second well is illustrated in
  Fig.\ref{fig:c30} for the octupole deformation $\beta_{30}$.
  It can be seen that the assumption of harmonicity is satisfied in
 the reegion of excitation energies up to 2-2.5 MeV.
 The root-mean-square deviation of the fit for $^{234}$U equals 1.3
  keV and 0.45 keV at the first and second minimum, respectively.
  Errors of the fit, for all nuclei and  deformations, fall in the range
  1-20 keV, with a typical number of points in each fit equal to 30.

\begin{figure}[h]
\vspace{-2mm}
\begin{minipage}[h]{80mm}
\centerline{\includegraphics[scale=0.8]{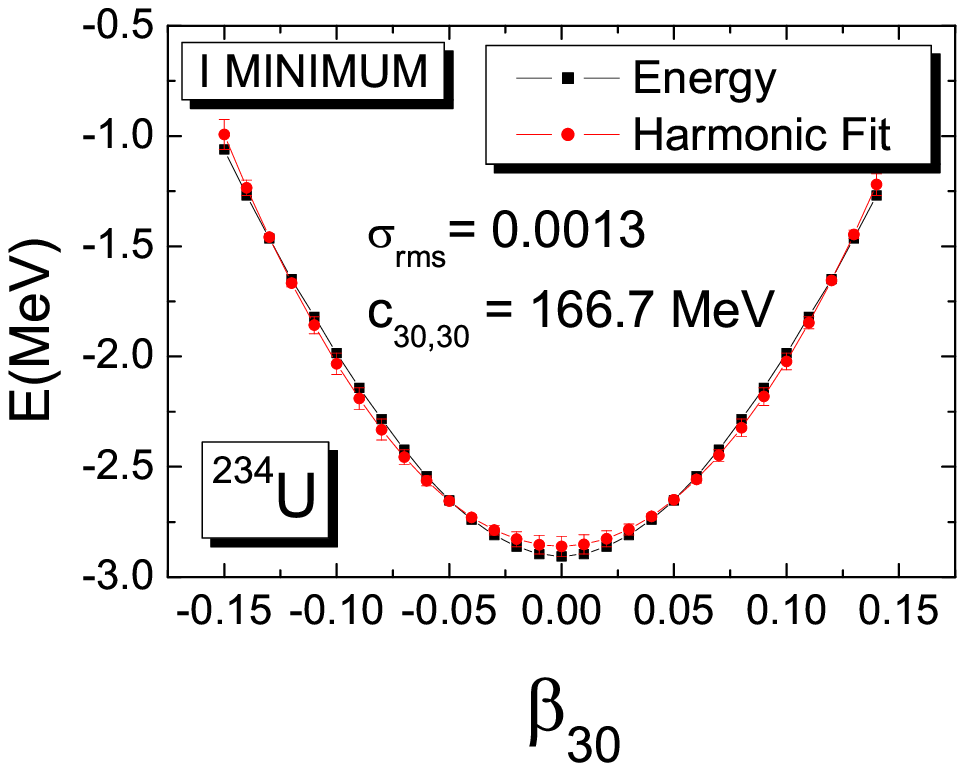}}
\end{minipage}
\begin{minipage}[h]{+80mm}
\centerline{\includegraphics[scale=0.8]{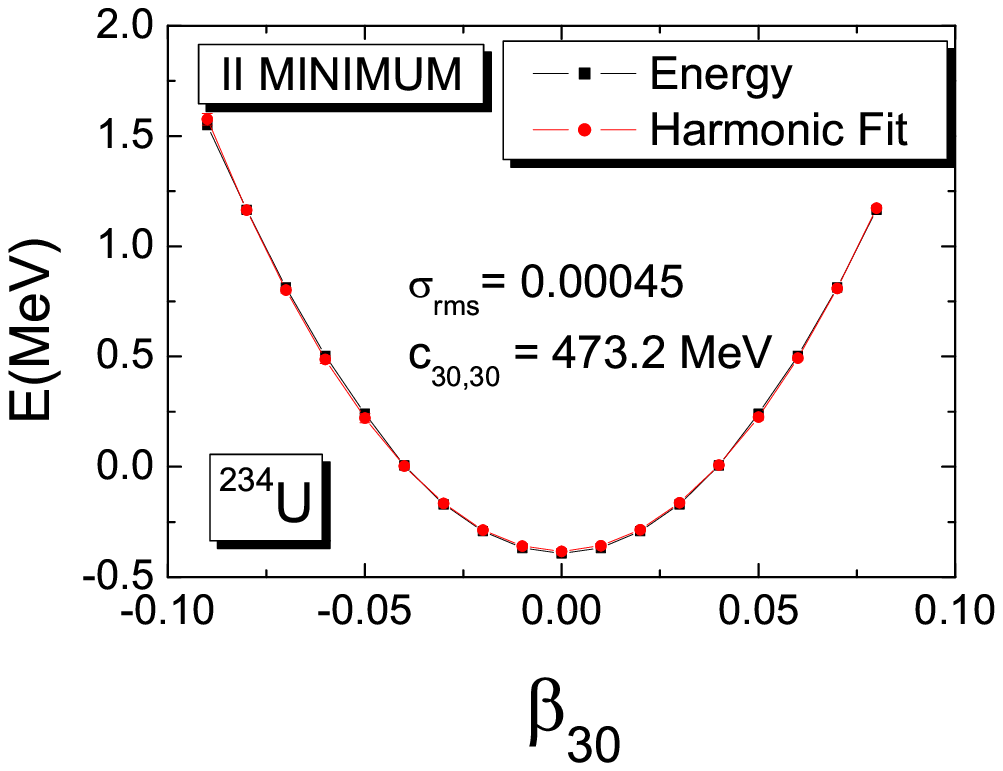}}
\end{minipage}
\caption{{\protect The extraction of the stiffness $C_{3030}$ at the I-st
and II-nd minimum for $^{234}$U.
 \label{fig:c30}
  }}
\end{figure}

 A collection of all calculated $K=0$ stiffness coefficients is shown in Fig.
\ref{fig:szt1} for the first minima and in Fig. \ref{fig:szt2} for the
 second minima. There are diagonal $C_{3030},C_{5050},C_{7070}$ and off-
diagonal $C_{3050},C_{3070},C_{5070}$ stiffness parameters. It is natural when
  the latter are smaller than the former. Very telling is the
 relative smallness of the $C_{3030}$ coefficients at the first minima,
 where they are similar in magnitude to the off-diagonal coefficients, and
  their much larger values at the second minima.

\begin{figure}[h]
\vspace{-3mm}
\begin{minipage}[l]{0mm}
\centerline{\includegraphics[scale=0.95]{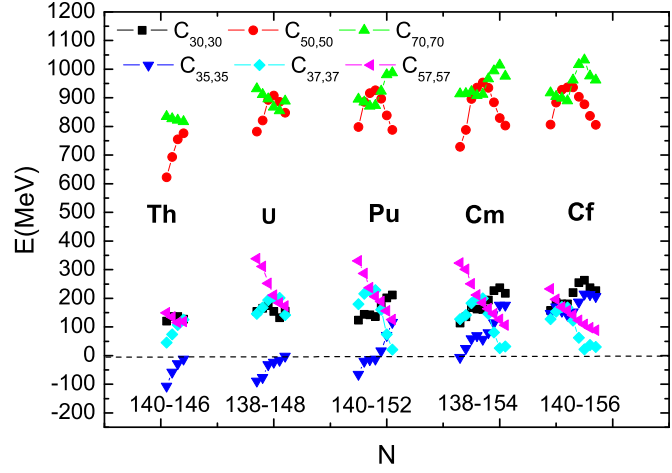}}
\end{minipage}
\caption{{\protect Stiffness coefficients at the I-st minimum.
  }}
  \label{fig:szt1}
\end{figure}

\begin{figure}[h]
\vspace{-3mm}
\begin{minipage}[l]{0mm}
\centerline{\includegraphics[scale=0.95]{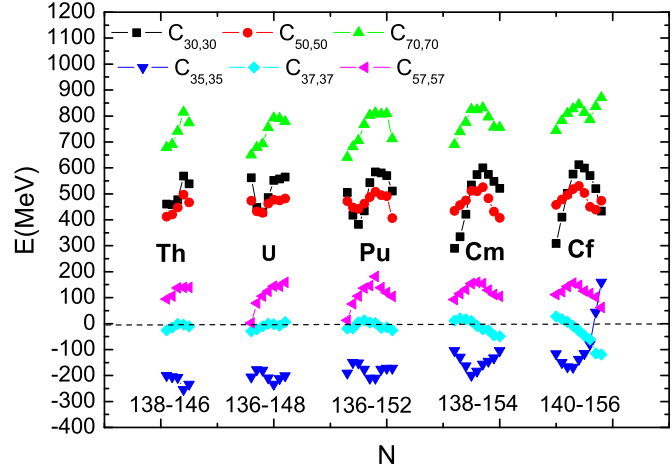}}
\end{minipage}
\caption{{\protect Stiffness coefficients at the II-nd minimum.
  }}
  \label{fig:szt2}
\end{figure}

 The plots of mass parameters $B_{3030}$ and $B_{5050}$ around the reflection-
 symmetric I and II minimum are shown in Fig. \ref{fig:b30b50}. The $^{240}$Pu
 nucleus has been chosen as an example. One can see that the assumption of
constant mass parameters, fixed at their values at the minima, is not
  drastically wrong. There are two reasons why a deformation dependence of
  mass parameters should not be crucial: 1) The relevant deformation range
      is limited, as the peak of the probability density of the one-phonon
 state corresponds to $\beta_{\lambda K}^{\pi-} \leq 0.1$;
  2) Phonon energy depends
 on the mass parameter as $(C/B)^{1/2}$, which is not a very
  strong dependence.
  One can observe a different behavior of $B_{3030}$ around the I-st and II-nd
  minima. While at the first minimum this parameter takes the maximal
  (or close to the maximal) value, it reaches the minimal value at the
  second minimum. Both $B_{3030}$ and  $B_{5050}$ are skewed with respect to the
 $\beta_{30}$ and $\beta_{50}$ axes.

\begin{figure}[h]
\vspace{-5mm}
\begin{minipage}[h]{80mm}
\centerline{\includegraphics[scale=0.85]{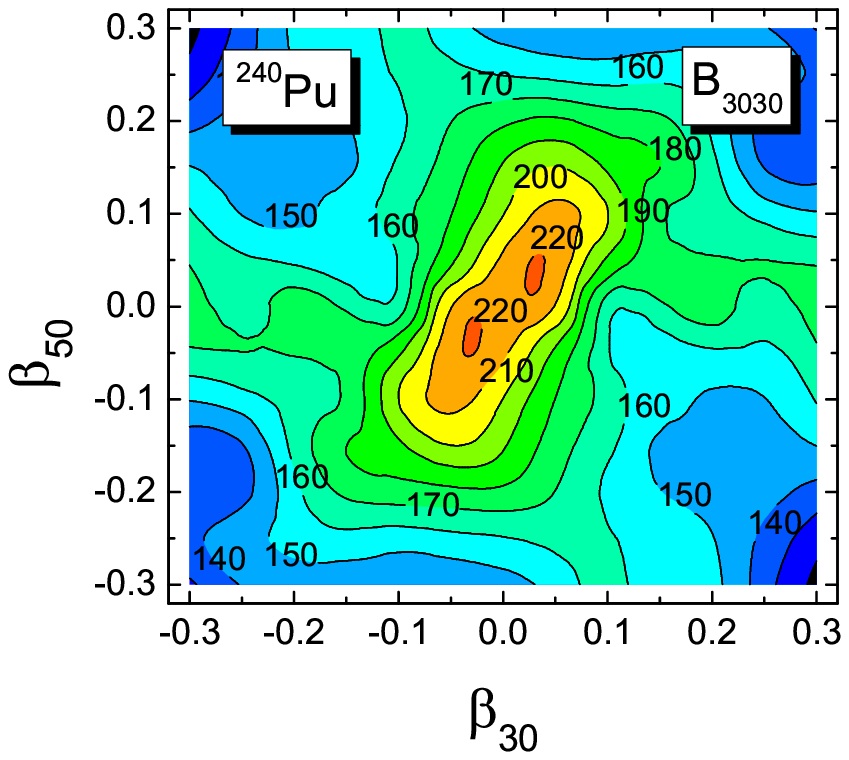}}
\centerline{\includegraphics[scale=0.85]{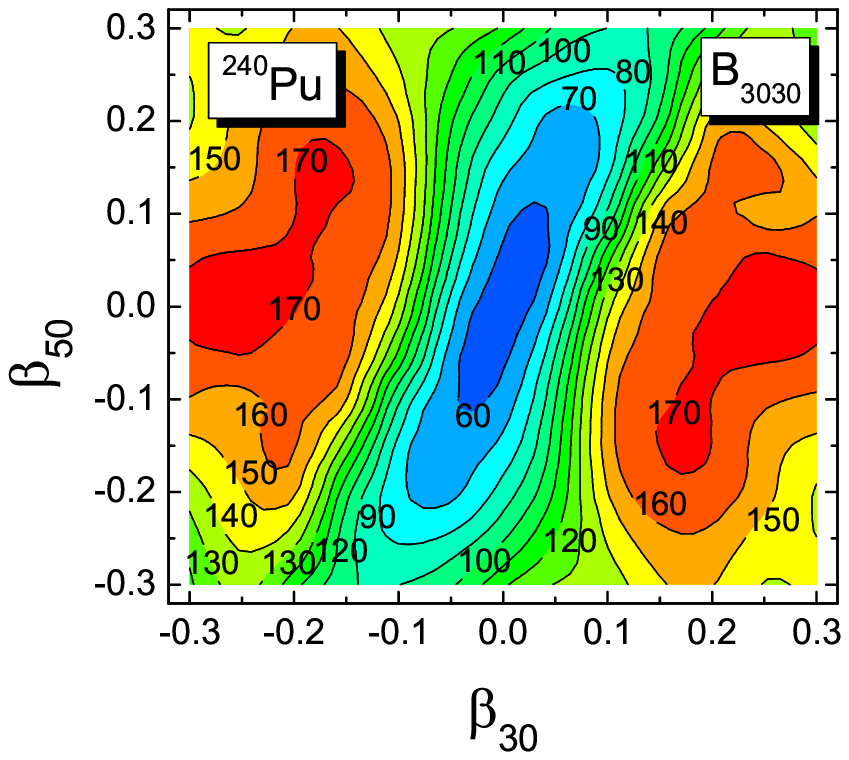}}
\end{minipage}
\begin{minipage}[h]{+80mm}
\centerline{\includegraphics[scale=0.85]{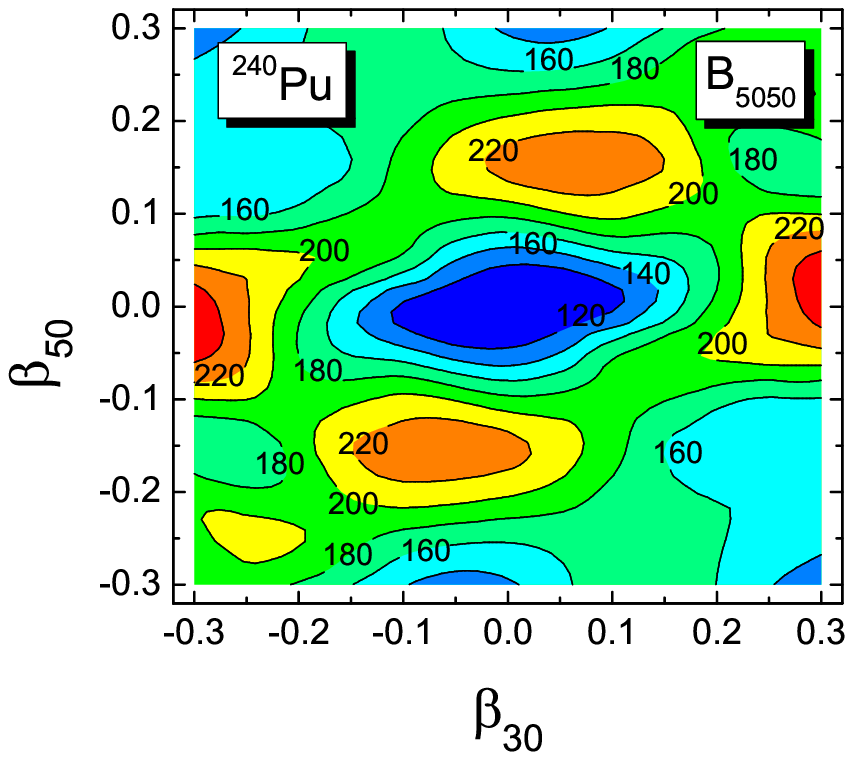}}
\centerline{\includegraphics[scale=0.85]{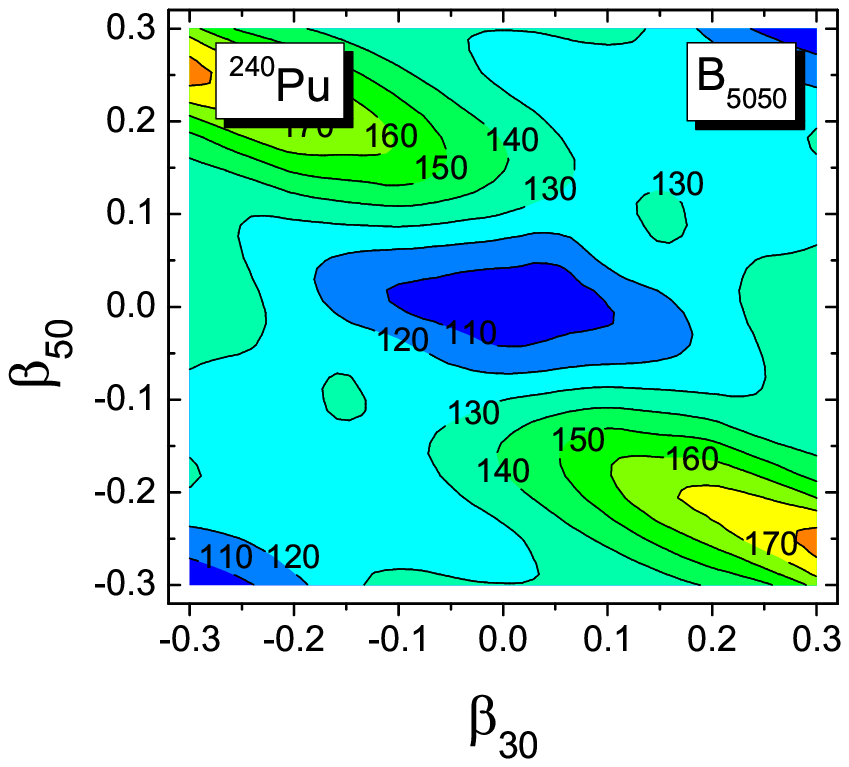}}
\end{minipage}
\caption{{\protect Cranking mass parameters $B_{3030},B_{5050}$ around
the first (top panels) and second (bottom panels) minimum for
$^{240}$Pu in the ( $\beta_{30},\beta_{50}$) plane.
 \label{fig:b30b50}
  }}
\end{figure}
\begin{figure}[h]
\vspace{-5mm}
\begin{minipage}[l]{0mm}
\centerline{\includegraphics[scale=0.95]{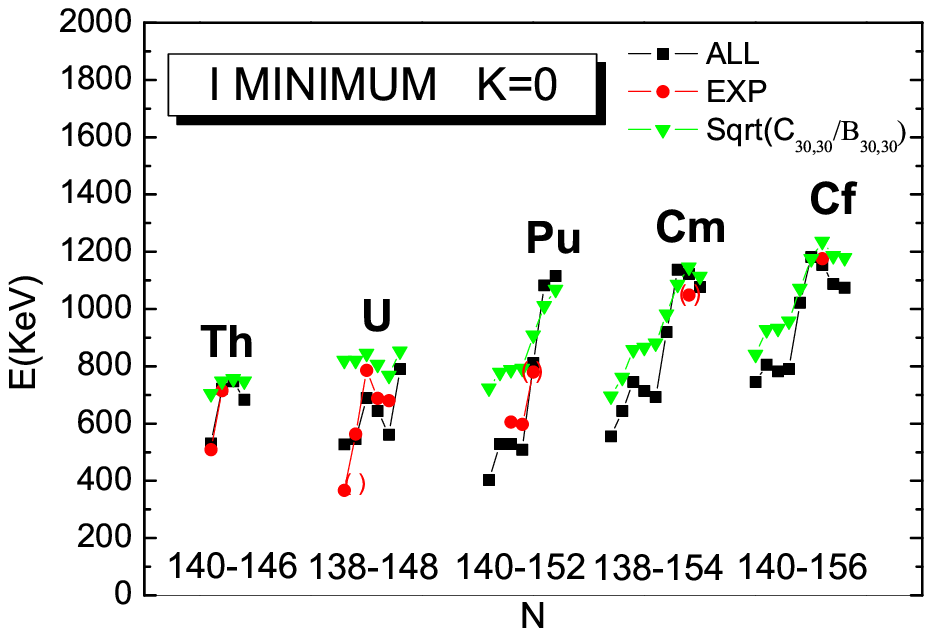}}
\centerline{\includegraphics[scale=0.95]{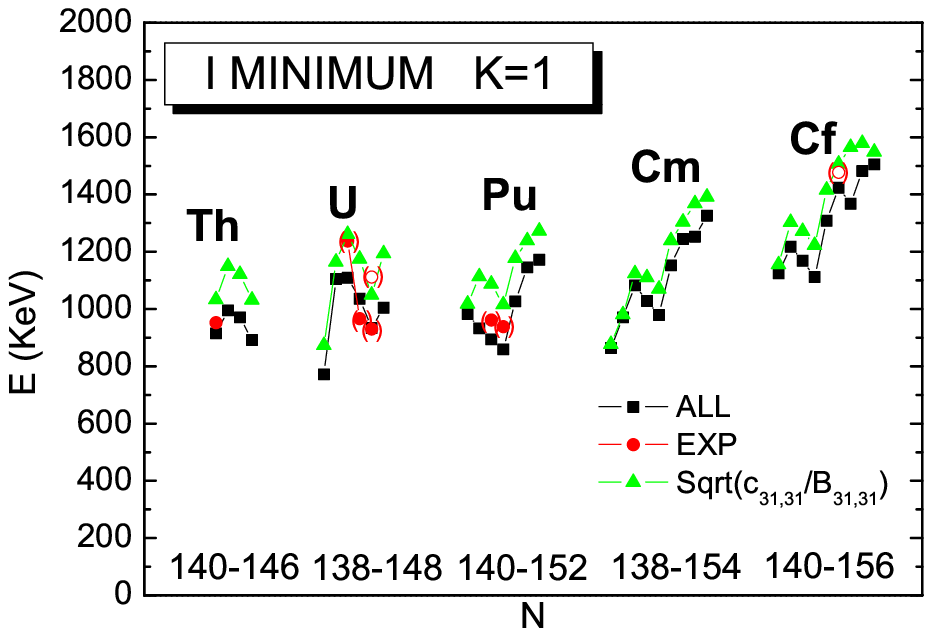}}
\centerline{\includegraphics[scale=0.95]{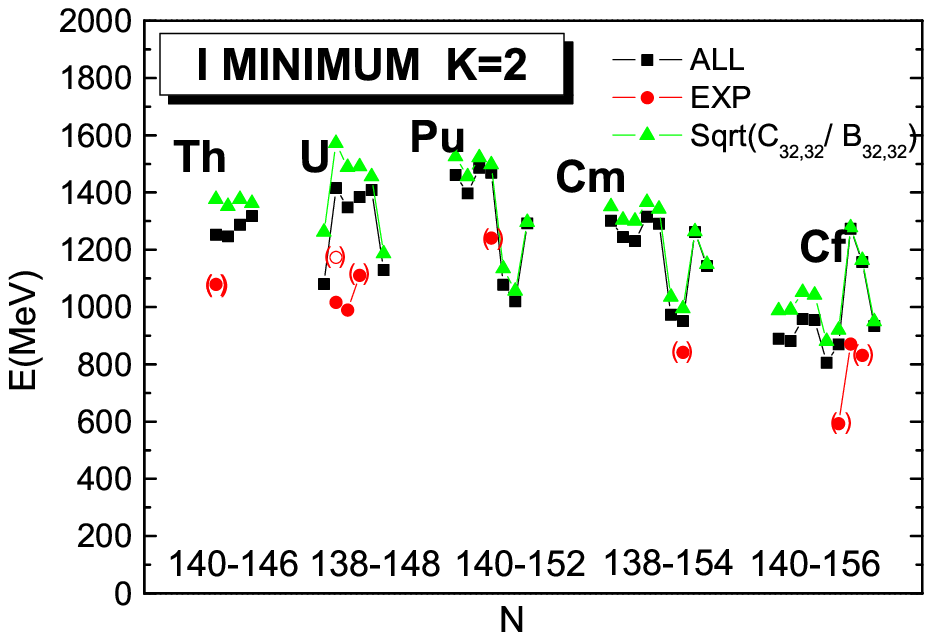}}
\end{minipage}
\caption{{\protect Energies of negative-parity shape oscillations in the g.s.
minimum for $K=$0,1,2. Results including all (ALL) multipolarities $\lambda=$
 3,5,7 are denoted by squares; triangles follow the
 formula: $\hbar \omega _{K} = \sqrt{\frac{C_{3K3K}}
{B_{3K3K}}}$. Experimental data (EXP) \cite{Firestone} are marked
with circles; parenthesis signals some uncertainty in the experimental
 assignment.}}
  \label{fig:vib1}
\end{figure}
 Results of the simpler, harmonic model (\ref{Hams}) that already contain the
 effect of the coupling of various multipolarities are presented below.
 Their modification due to the variability of the mass parameters was
  worked out for selected cases and is discussed afterwards.

  \subsection{Results with constant mass parameters}

 Calculated energies of the lowest negative-parity shape oscillations in the
 first minimum for various even-even actinides are shown in
 Fig. \ref{fig:vib1}
 together with experimental results \cite{Firestone}. Only vibrational
  cases are considered which means that we excluded nuclei with
  the octupole equilibrum deformation or very close to the reflection-symmetry
  breaking.
 As the calculated energies of the $K=3$ phonons are greater than 1.5 MeV
  in Th, U, Pu and Cm isotopes and greater than 1.3 MeV in Cf isotopes,
 and, moreover, an experimental information on such
 octupole states is scarce and  uncertain, we do not show these results here.
 It may be seen that for the $K=0$ and $K=1$ modes an overall agreement to
  within $\sim$100 keV (except for $^{230}$U) is obtained between the data and
 theoretical evaluation. It is worth mentioning that our calculations do not
 involve any adjustment to the data they are compared with.
 For the $K=2$ mode the experimental energies are considerably
 overestimated. On the other hand, one has to notice that the experimental
 assignments made for $K=1$ and $K=2$ octupole band-heads are not many and some
  of them seem quite uncertain.
  One can observe in Fig. \ref{fig:vib1} a clear change in energies between
 $N=146$ and $N=148(150)$: they {\it increase} for $K^{\pi}=0^-$ and $1^-$
 in U, Pu, Cm and Cf, and {\it decrease} for $K^{\pi}=2^-$ in U, Pu and Cm.

  From the difference between the energies and the quantities
 $(C_{3K3K}/B_{3K3K})^{1/2}$ one can judge the importance
 of the off-diagonal components of the stiffness and mass tensors. The
 resulting mixed multipolarity character of the lowest phonon may be
  also seen from the deformation parameters $\beta_{\lambda K}^{tr}$
  that follow from Eq.(\ref{coor}).
   For the $K=0$ mode, they show a small admixture of $\beta_{50}$ and/or
  $\beta_{70}$ deformation to $\beta_{30}$: $\lambda=5$ mixing is important
  for the lightest/heaviest isotopes, $\lambda=7$ for the middle.
  For example, we have $(\beta_{30}^{tr},\beta_{50}^{tr},\beta_{70}^{tr})=
 (0.049,0.013,-0.015)$ in $^{232}$ U, $(0.058,0.005,-0.017)$ in $^{238}$U and
  similar values in $^{240}$Pu and $^{246}$Cm, while
  the $\beta_{30}$-$\beta_{50}$ coupling  is implied by the values
 $(0.052,-0.017,0)$ in $^{250}$Cf.
 For the $K=1$ mode, the admixture of $\beta_{51}$ is, except for Th isotopes,
 larger than that of $\beta_{71}$: in $^{238}$U we have
  $\beta_{\lambda 1}^{tr}=(0.050,-0.016,0.007)$; similar values occur for
  $^{240}$Pu and many other nuclei.
 As shown in Fig.  \ref{fig:vib1}, the role of the coupling
 to higher multipolarities decreases with the increasing $K$-number in the
 first potential well. In particular, for $K^{\pi}=2^-$ vibrations, one could
 use the simple formula $\hbar \omega _{2} =
 \sqrt{\frac{C_{3232}}{B_{3232}}}$ for the excitation energy.
 Appreciable admixtures, mainly of $\lambda=5$, are present for the
 lighter isotopes.

In Fig. \ref{fig:vib2}, calculated low-lying negative-parity excitation
energies at the second minimum are shown as a function of the neutron number
for different isotopes. The plotted
experimental data were taken from \cite{Thirolf}.
 The $K=3$ energies are not presented as they lie above 2.5 MeV,
 even higher than in the first minima.
 We do show calculated vibration energies in the second well in Cf isotopes,
  although no shape isomer was established there up to now.
  In our calculations, the barrier protecting the second minimum in $^{246}$Cf
  against fission is by 1 MeV smaller and thinner than in its isotone
  $^{244}$Cm, which supports the short-lived ($< 5$ ps) shape isomer and its
  longer lived ($< 100$ ns), probably spin-isomeric, 1.3 MeV excitation
  \cite{sletten}.
  However, it is not completely excluded that, with a better technique,
  some shape isomer in Cf could be established in the future, at
  least in an odd isotope.

Rather extensive experimental results were collected on the
excitations in the II minimum of $^{240}$Pu \cite{240PuIIa,Gassmann}.
 These include negative-parity rotational
  bands as well as the candidate for the $\beta$-vibration band.
  The lowest rotational bands interpreted as $K^{\pi} = 0^{-}, 1^{-}$ and
  $2^{-}$ have bandheads at the excitation energy 555 ($I^{\pi}=1^{-}$), 836
   ($I^{\pi}=1^{-}$) and 806 keV ($I^{\pi}=2^{-}$), respectively.
   Only two first states, $I^{\pi}=1^{-}$ and $3^{-}$, of the
  supposed octupole $K^{\pi}=0^{-}$ band were observed. There are three
  more bands identified as $K^{\pi}=1^{-}$: at 936, 1246 and 1344 keV;
 the supposed $\beta$ band starts with the second $I^{\pi}=0^{+}$ state
 at 770 keV.

As follows form Fig.\ref{fig:vib2},
 within the presented model we cannot reproduce the $K=0^-$ excitations
  at the second minimum.
 In $^{240}$Pu, we obtain energy nearly three times too large.
Situation is even worse for $^{236}$U, in which the
back-decay to the first minimum was experimentally detected and
 the excitation energy of mere 290 keV was attributed to the
 $K^{\pi}=0^{-}$ state in the second well.
 One reason for our results can be seen in
 Figs. \ref{fig:szt1} and \ref{fig:szt2}, by comparing the
 stiffness $C_{3030}$ against the octupole deformation $\beta_{30}$
 at the first and second minima. For all studied nuclei, the
 stiffness in the second minimum is roughly two times larger than in
 the first one and mostly larger than the stiffnes $C_{5050}$ against
 $\beta_{50}$.
 The second reason for the larger $K=0$ excitation energies in the second well
 comes from the mass parameters: $B_{3030}$ are two to three times smaller
 at the second minimum than at the g.s., and even little smaller than
  $B_{5050}$, opposite to the situation at the first minimum,
 see Fig. \ref{fig:b30b50}.

 As a result, the lowest $K^{\pi}=0^-$ excitation in the second well is a
 mixture of the $\beta_{30}$ and $\beta_{50}$ modes, with the latter dominating
  for nearly all isotopes. For example, the values of $\beta_{\lambda 0}^{tr}$
 are equal to $(0.033,0.040,0)$ in $^{230}$Th, $(0.032,0.041,-0.005)$ in
 $^{236}$U,
 $(0.030,0.040,-0.007)$ in $^{240}$Pu, $(0.024,0.041,0)$ in $^{248}$Cm and
 $(0.020,0.040,0.003)$ in $^{250}$Cf.
 When the coupling between different multipolarities is neglected,
  the $\beta_{50}$ mode is the lowest in the second minimum.

 There is a large energy lowering below
  the uncoupled value $(C_{3030}/B_{3030})^{1/2}$ due to this coupling,
  see Fig. \ref{fig:vib2}.
  This shows that it is impossible to describe low-lying $K^{\pi}=0^-$
  states in the second well as a pure octupole vibration.
  However, the coupling with the $Y_{50}$ mode is not sufficient in view of
  the experimental results.

  The couplings for the $K^{\pi}=1^-$ and $2^-$ modes reduce phonon energies by
 $\sim 300$ and $\sim 400$ keV, respectively, below the values of
 $(C_{3K3K}/B_{3K3K})^{1/2}$, Fig. \ref{fig:vib2}.
 This is much less than for the $0^-$ mode. The values
  $\beta^{tr}_{\lambda K}$ show a conspicuous mixing of $\lambda=3$ and 5,
  with much smaller admixtures of $\lambda=7$. For $K^{\pi}=1^-$ phonons, the
  octupole mode always dominates, $\beta^{tr}_{31} > \beta^{tr}_{51}$.
   In most cases this is also true for the $K^{\pi}=2^-$ mode, but for
  some nuclei $\beta^{tr}_{52}$ reaches the size of $\beta^{tr}_{32}$.
 Both phonon energies in $^{240}$Pu are overestimated,
 by 400 ($1^-$) and 170 keV ($2^-$); the supposed $2^-$ band head at 830 keV
 in $^{236}$U is also calculated too high by 430 keV. The difference in the
 calculated $K^{\pi}=2^-$ phonon energies in $^{236}$U and $^{240}$Pu follows
 mainly from the larger mass parameter $B_{3232}$ in the latter.
 It is remarkable that in the second minimum the mass parameters
 $B_{3131}>B_{3232}$ are $\sim$2 or more times larger than $B_{3030}$, while
 the latter are the largest in the first minimum.

  \subsection{Results with shape-dependent mass parameters}

   Oscillation energies $\hbar\omega_K$ in the first and second minima
  calculated with shape-dependent mass parameters
  for selected even-even actinides are compared
  in Table \ref{1}  to the constant-mass variant of the calculations
  and the experimental data, where available.
  As may be seen, the model including the shape dependence of the mass
  parameters (\ref{Ham}) spoils or improves the agreement with the data,
 depending on a particular case. In the first well, there is a substantial
  decrease in some $K^{\pi}=0^-$ energies away from the measured values,
 especially for $^{240}$Pu. Some other energies become closer to the data,
   e.g. in
    $^{236}$U and $^{246}$Cm. The $K^{\pi}=1^-$ energy in $^{230}$Th is
  spoiled, while for the $K^{\pi}=2^-$ mode there is an improvement for
  $^{240}$Pu and worsening for $^{250}$Cf, already 400 keV off the data in
   the harmonic approximation.
  These changes follow from the differences between the mass parameters
  at the minimum and their averages over a region around it.

   In the second minimum, $K^{\pi}=0^-$ energies from the full Hamiltonian
  (\ref{Ham}) are by 200-300 keV lower than for its simplified version
  (\ref{Hams}), except for $^{234}$U, for which $\hbar\omega_0$ stays the same.
   This is an improvement, but not a sufficient one, as now
 the $K^{\pi}=0^-$ energy for $^{240}$Pu is roughly 2.4 times too large
  instead of three.
  The $K^{\pi}=1^-$ and $2^-$ energies in $^{236}$U and $^{240}$Pu
   do not change
   much, except for the $K^{\pi}=2^-$ state in $^{240}$Pu, that increases
   further 200 keV away from the experimental value.

  The structure of vibration phonons changes mainly according to the energy
  change. The values of $\beta_{\lambda K}^{tr}$ increase appreciably with
  respect to the constant mass parameters version only if there is
  an appreciable decrease in the phonon energy.
     Their ratios are similar as those for the harmonic Hamiltonian.
  It should be noticed that for anharmonic vibrations the simple relation
 between $\beta_{\lambda K}^{tr}$ and the multipolarity
 composition of the vibration coordinate, based on Eq.(\ref{coor}), is lost.

   One can say that the results obtained by the full diagonalization of
  the Hamiltonian (\ref{Ham}) do not introduce drastic changes in the
    results obtained with (\ref{Hams}).

\begin{figure}[h]
\vspace{-5mm}
\begin{minipage}[l]{0mm}
\centerline{\includegraphics[scale=0.95]{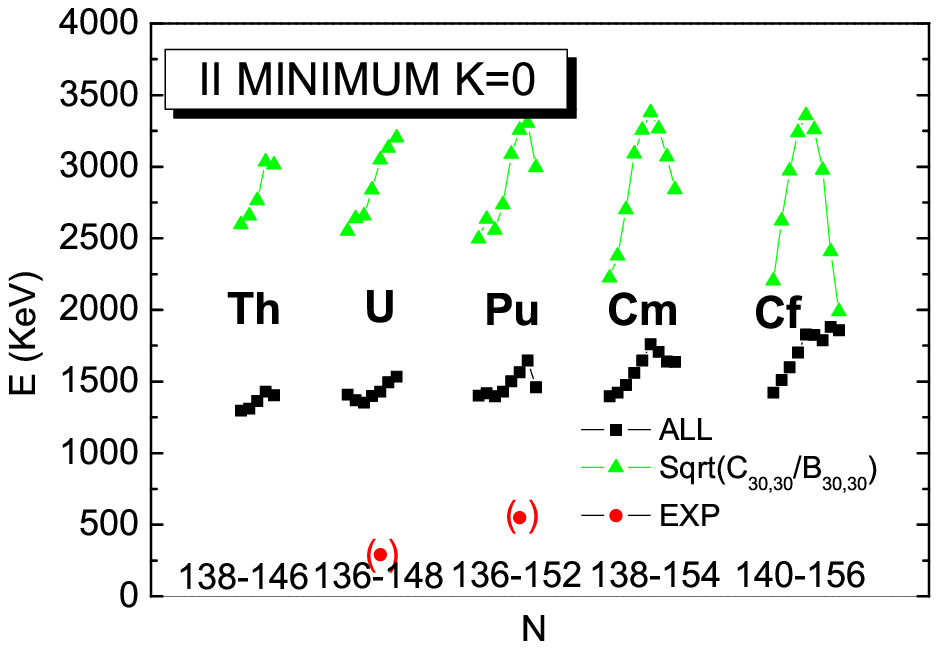}}
\centerline{\includegraphics[scale=0.95]{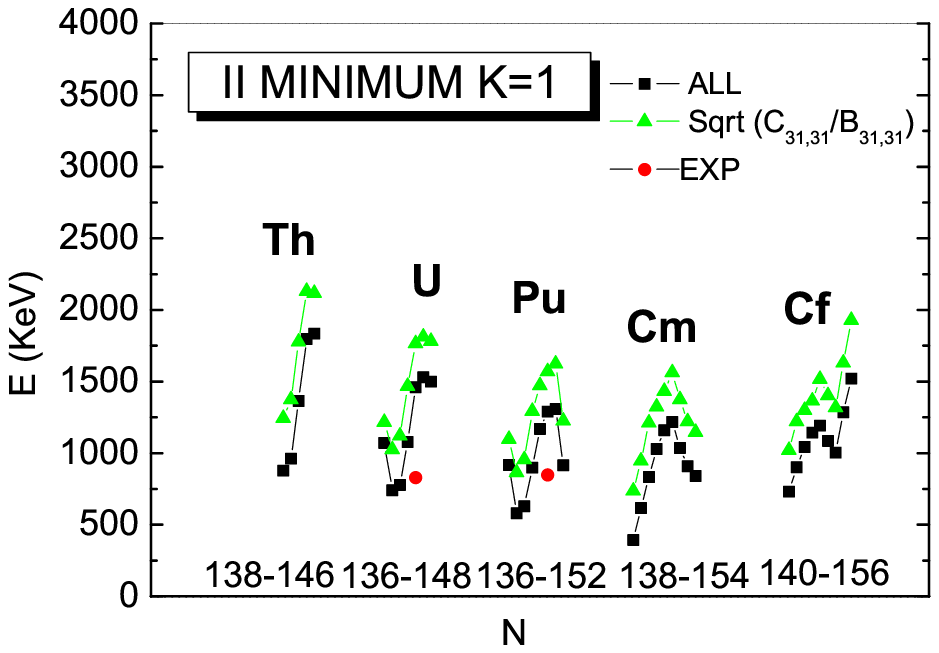}}
\centerline{\includegraphics[scale=0.95]{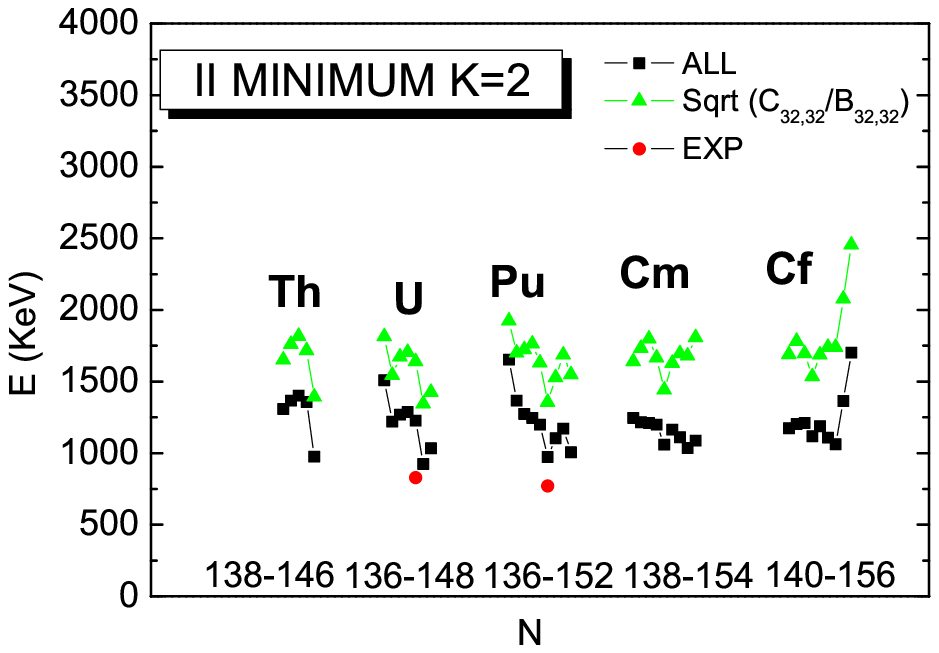}}
\end{minipage}
\caption{{\protect The same as in Fig \ref{fig:vib1},
 but in the second minimum. Experimental data (EXP)
are taken from \cite{Thirolf}.
  }}
  \label{fig:vib2}
\end{figure}

\begin{table}
\caption{\label{1} Energies of negative-parity shape vibrations (in keV) for
 $K=0,1,2$ around I and II minimum obtained from the diagonalization of the
 full Hamiltonian (\ref{Ham}) - $H_{coll}$, using constant mass
parameters in the minimum (\ref{Hams}) - $B_{const}$ and
calculated by a simple formula: $\hbar \omega _{K} =
\sqrt{\frac{C_{3K3K}}{B_{3K3K}}}$. Experimental data
   are taken from \cite{Firestone} for the I-st minima and from
  \cite{Thirolf} for the II-nd minima.}
\begin{ruledtabular}
\begin{tabular}{cc|cccc|ccccc}

  &  & &  & I MIN &  & & &   II MIN & \\

   $Z$ &   $A$ &   $H_{coll}$ &  $B_{const}$ & $\sqrt{\frac{C_{3K3K}}{B_{3K3K}}}$& $exp$ &$H_{coll}$ & $B_{const}$ &$\sqrt{\frac{C_{3K3K}}{B_{3K3K}}}$ & $exp$ \\
\hline
  &  &$K=0$ &  &  &  &$K=0$& & & \\
\hline
92&    230 &    551 &   527&   756&   367 & 1138&  1309&  2482& $-$ \\
92&    232 &    698 &   546&   773&   563 & 1132&  1331&  2430& $-$\\
92&    234 &    820 &   689&   906&   786 & 1400&  1390&  2672& $-$\\
92&    236 &    694 &   644&   881&   688 & 1199&  1430&  3002& $290$\\
92&    238 &    451 &   561&   795&   680 & 1236&  1492&  3104& $-$\\
94&    238 &    465 &   529&   814&   605 & 1197&  1505&  2999& $-$\\
94&    240 &    338 &   509&   792&   597 & 1272&  1565&  3254& $555$\\
94&    244 &    1062&   1083&  1129&  950(?)   & 1353&   1654&  3339& $-$\\
96&    246 &    1092&   1138&  1217&  1079& 1380&   1681&  3440& $-$\\
\hline
  &  &$K=1$ &  &  &  &$K=1$& & & \\
\hline
90&    230 &    770 &   915&  1033&   952 & $-$ & 962 & 1373& $-$ \\
92&    236 &    $-$ &  1036&   1175&  967(?) & 1354& 1461& 1765& $-$ \\
94&    240 &    858 &   859&  1015&   938(?) & 1266&  1289& 1568& $836$\\
\hline
  &  &$K=2$ &  &  &  &$K=2$& & & \\
\hline
90&    230 &   1264 &  1253&  1376&   1079(?) & $-$ & 1367& 1760& $-$ \\
92&    234 &   1377 &  1348&  1488&  989 & $-$ & 1287& 1704& $-$\\
92&    236 &    $-$ &  1385&   1490& 1110(?) & 1345& 1226&1639& $830$(?)\\
94&    240 &   1295 &  1469&  1497&   1241(?)& 1166&   972& 1355& $806$\\
98&    250 &   1303 &  1274&  1277&   872 & $-$ & 1061& 1738& $-$\\
\end{tabular}
\end{ruledtabular}
\end{table}
\endgroup

 \subsection{Electromagnetic dipole transitions}

 Transition dipole moments governing E1 transitions between one-phonon $K=0$
 and $K=1$ bands and the g.s. band have been calculated using the values
 $\beta_{\lambda K}^{tr}$ from the model (\ref{Hams}) with constatnt mass
 parameters. They are given in Table \ref{2} for selected nuclei.
 At the first minimum, $D^t$ values show a considerable variation from
 a nucleus to nucleus. The $K=0$ and 1 components are correlated, e.g.
 both are large for $^{230}$U; this large $K=0$ value is roughly consistent
 with the measurements \cite{230U}. For $K=1$, $B(E1)\sim 2(D^{t})^2$,
 Eq.~(\ref{Btran}), so for an easy comparison of
 $B(E1)$s for $K=0$ and $K=1$, one should multiply $D^t$ for $K=1$
 from Table \ref{2} by $\sqrt{2}$.
   For some nuclei, the opposite signs of the microscopic and macroscopic
 parts lead to a nearly complete cancellation of $D^t$.
 This results in a variation of the deexcitation pattern along the isotopic
 chain, as may be seen for $^{230-238}$U.

 More specifically, the negative-parity rotational band decays according
 to the ratios of the rates of E2 transitions within, and E1 transitions
  out of the band. From the calculated transition dipole moments $D^t$ in
 [$e$ fm] we can estimate the ratio
 \begin{equation}
  \label{ratio}
\frac{T(E1)_{I\rightarrow I-1}}{T(E2)_{I\rightarrow I-2}}= 1.303
 \frac{E_{\gamma}^3(E1)}{E_{\gamma}^5(E2)}\left(\frac{8}{5}\right)
 \frac{(2I-1)(I-1)} {(I-1)^2-K^2}\left(\frac{D^t}{Q_0}\right)^2  ,
  \end{equation}
 where $Q_0$ is the intrinsic quadrupole moment of the negative-parity band
  in units of [10 b], and energies $E_{\gamma}$ of gamma transitions are
  given in MeV.
 Using values of $Q_0$ measured for the g.s. bands, we obtain the following
 T(E1)/T(E2) ratios for transitions from the $I^{\pi}=7^-$ state of the
 $K^{\pi}=0^-$ band:
  around 300 in $^{232,234}$U, 1.6 in $^{236}$U and around 100 in
  $^{238}$U. Although not a perfect match with tha data, this roughly
 correlates
 with no intraband E2s seen along the negative-parity bands in $^{232,234}$U,
the complete regular E2 cascade and no E1s above the $3^-$ state in $^{236}$U,
  and the E2 cascade ending at the $7^-$ state, with more E1s in $^{238}$U
 \cite{Zeyen,Browne,238U}. On the other hand, the predicted nearly perfect
 cancellation of $D^t$ for $K=0$ in $^{240}$Pu is not supported
 by the value $\approx$0.12 $e$fm measured at spin 11 and the {\it smaller}
  values for the neighbouring Pu isotopes \cite{240Pu}.
  The relatively small dipole moment measured for $^{230}$Th \cite{230Th}
  does not contradict our $K=1$ value.

  In the isomeric second minimum the situation is entirely different.
  The macroscopic part, $D_{mac} \sim \beta_{20} \beta_{3K}$ in the leading
  order, is dominant there as its magnitude becomes larger than that of
  the shell-correction part due to the large equilibrium
  value of $\beta_{20}$. For $K=0$ bands,
 the microscopic part adds to the macroscopic part.
  The large values of the dipole moment and small energies of rotational
  transitions near the bandheads preclude observation of the intraband E2s
 in negative-parity vibrational bands in the isomeric minima in actinides.
 This agrees with the lack of experimentally observed
  E2 transitions in the isomeric minimum in $^{240}$Pu \cite{Thirolf,240PuIIa,
 Gassmann}.

\begin{table}
\caption{\label{2} Macroscopic $D_{mac}$, microscopic
$D_{mic}$ and total transition dipole moment $D^t$ for $K^{\pi}=0^-$ and $1^-$
  phonons in the first and second minimum.}
\begin{ruledtabular}
\begin{tabular}{cc|ccc|cccc}
 &  &  & I MIN & &  & II MIN & \\
   $Z$ &   $A$ &   $D_{mac}$ &  $D_{mic}$ & $D^t$ &   $D_{mac}$ & $D_{mic}$ & $D^t$ \\
\hline
  &     & $K=0$    &  &     & $K=0$    &  &  \\
\hline
92&    230 &    0.19  &   0.04 &   0.23 &   0.29 & 0.11 & 0.40 \\
92&    232 &    0.17  &   0.00 &   0.17 &   0.34 & 0.08 & 0.42 \\
92&    234 &    0.17  &   -0.07 &  0.10 &   0.34 & 0.04 & 0.38 \\
92&    236 &    0.17  &   -0.16 &  0.01 &   0.34 & 0.02 & 0.35 \\
92&    238 &    0.17  &   -0.24 & -0.07 &   0.33 & 0.02 & 0.35 \\
94&    238 &    0.19  &   -0.13 &  0.06 &   0.33 & 0.09 & 0.42 \\
94&    240 &    0.18  &   -0.18 &  0.00 &   0.32 & 0.08 & 0.40 \\
94&    244 &    0.15 &   -0.18 &   -0.03 &  0.29 & 0.03 & 0.32\\
96&    246 &    0.15 &   -0.14 &    0.01 &  0.29 & 0.12 & 0.41 \\
\hline
  &     &$K=1$&  &     & $K=1$  &  &  \\
\hline
90&    230 &   -0.042 &   0.031&  -0.01 &  -0.157&-0.077&-0.23 \\
92&    230 &   -0.049 &  -0.079&  -0.13 &  -0.173&-0.055&-0.23 \\
92&    232 &   -0.047 &  -0.041&  -0.09 &  -0.179&-0.035&-0.21 \\
92&    234 &   -0.049 &    0.008& -0.04 &  -0.174&-0.017&-0.19 \\
92&    236 &   -0.055 &    0.035& -0.02 &  -0.170& 0.00 &-0.17 \\
92&    238 &   -0.063 &    0.054& -0.01 &  -0.174& 0.00 &-0.17 \\
94&    238 &   -0.059 &    0.024& -0.036&  -0.182& 0.023&-0.16 \\
94&    240 &   -0.065 &    0.045& -0.02 &  -0.183& 0.022&-0.16 \\
\end{tabular}
\end{ruledtabular}
\end{table}


\section{Conclusions}

   Our study of low-lying negative-parity oscillations in even-even actinides
  leads to the following conclusions:

\begin{itemize}

      \item {Considering that we have no adjustable parameters, the data on
           negative-parity excitations in the first minima
             are reproduced quite well for the $K^{\pi}=0^{-}$
          and $1^-$ phonons; energies of the $2^{-}$ phonons are significantly
           overestimated.}

      \item {The model predicts $K^{\pi}=0^{-}$ energies in the second well
             in $^{240}$Pu and $^{236}$U that are three and more times larger
             than the claimed experimental values.
            For the $K=$1,2 phonons, the agreement with the data is
          better, similar to that obtained in \cite{RPAR} for $^{240}$Pu,
        but still the calculated energies are too high.}

      \item {Low-lying negative-parity oscillations show small admixtures
            of the multipolarities $\lambda=5$,7 to the octupole mode
            at the first minima, and equal or dominant contribution of the
            $\lambda=5$ multipole at the second minimum. Hence the
             coupling of various multipolarities is important
             in the description of the "octupole" vibrations, especially
            in the second minimum.}

      \item {Taking into account the multipoles $\lambda=$3,5,7
            in the phonon structure and the exact macroscopic contribution
          to the dipole moment, we predict large transition dipole moments
            from the "octupole" band to the g.s. band
             in lighter actinides at the first minima, and for all
            investigated nuclei at the isomeric minima.}

\end{itemize}

       We do not see any way to reconcile our model with the reported data
       on the energies of the $K^{\pi}=0^-$ mode in the second well.
       Both the calculated sizable stiffness Fig. \ref{fig:szt2} and small
       mass parameters Fig. \ref{fig:b30b50} suggest that
       either our model is completely unreliable there, or, perhaps, the
       experimental $K^{\pi}=0^-$ assignments in the shape isomeric
       minima in $^{236}$U and $^{240}$Pu are not related to the collective
       shape vibrations.

      One could think of possible improvements of the model. A natural
     step would be to include the quadrupole-octupole coupling.
      However, as long as the second minima are reflection-symmetric,
      this would be a second-order effect, while the energy surfaces
       and mass parameters do not hint to its unusual enhancement.
       One could also consider a fine-tuning of the pairing strength
       in the second well, or including the quadrupole pairing as
    in \cite{RPAR}. Still, the required reduction
       of the $0^-$ phonon energies is so large, that the pairing alone
       hardly can be a cause.
       On the other hand, the discrepancies observed for $1^-$ and $2^-$
       phonons in the II-nd well probably could be reduced by a fine tuning of
       the model parameters.

\appendix
\section{Some aspects of the calculations}

  The matrix diagonalization of the Hamiltonian (\ref{Ham}) is performed
  in the basis of the three-dimensional harmonic oscillator.
   The matrix elements of the Hamiltonian were calculated with the
  20 point Gauss-Hermitte quadratures. The calculations were reduced by
  setting the mass parameters to their values at the minimum outside
  the mesh of $14\times 14 \times 14$ points for the easier $K=0$ case,
  and outside the mesh of $8\times 8 \times 6$ points for $K=1$ and 2.
   We have checked that the related distortion of the Hamiltonian is
 unimportant for the lowest phonon states, as even
  the smaller mesh covers their peak region.
  As a check of the program, we reproduced the harmonic case of
  constant mass parameters. We also checked the case of the position-dependent
  mass parameter $B \sim \beta^2$ with the quartic potential $\sim \beta^4$,
  which gives the harmonic spectrum.

  The cranking mass parameters were calculated by replacing derivatives
  with respect to deformations by finite differences. This has some
   advantages for many deformation parameters, but introduces an error which
  may reach 2-3\% for diagonal components. The new approach was tested
  with the old code for mass parameters for axial and some non-axial
  deformations.
  The main tool of these calculations, the code that diagonalizes the
  s.p. Hamiltonian, was checked independently with the older, less general
   versions.

  An independent test of this code is provided by the fact, that different
 deformation sets may correspond to the same shape, so they should
  produce the same s.p. spectrum. In particular, the deformations
   $Y_{\lambda K s}$ with $K=1,3$, $\lambda=3$,5,7 of an  axially
  symmetric equilibrium shape give the same s.p. spectrum as the deformations
   $Y_{\lambda K c}$. The latter are accomodated by the parametrization
   Eq. (\ref{shape}), by switching the choice of the symmetry axis from
   $z$ to $y$.
   This follows from the relations between spherical harmonics
  defined with respect to the reference axes $z,x,y$ (denoted as $Y^y$) and
  the standard ones:
  \begin{eqnarray}
   Y^y_{3 1 c} & = & \sqrt{\frac{3}{8}} Y_{3 0}+\sqrt{\frac{5}{8}} Y_{3 2 c}  , \\ \nonumber
   Y^y_{3 3 c} & = & -\sqrt{\frac{5}{8}} Y_{3 0}+\sqrt{\frac{3}{8}} Y_{3 2 c}  ,  \\ \nonumber
   Y^y_{5 1 c} & = & -\sqrt{\frac{15}{64}} Y_{5 0}-\sqrt{\frac{7}{16}} Y_{5 2 c} -\sqrt{\frac{21}{64}}  Y_{5 4 c}  , \\ \nonumber
   Y^y_{5 3 c} & = &  \sqrt{\frac{35}{128}} Y_{5 0}+\sqrt{\frac{3}{32}}
   Y_{5 2 c} -\frac{9}{\sqrt{128}} Y_{5 4 c}  , \\ \nonumber
   Y^y_{7 1 c} & = & \frac{5\sqrt{7}}{32} Y_{7 0}+\frac{15\sqrt{7}}{32\sqrt{2}}
    Y_{7 2 c} + \frac{3\sqrt{33}}{32} Y_{7 4 c} +
    \frac{\sqrt{429}}{32\sqrt{2}} Y_{7 6 c}  , \\ \nonumber
   Y^y_{7 3 c} & = & - \frac{3\sqrt{21}}{32} Y_{7 0} - \frac{19}{32\sqrt{2}}
   Y_{7 2 c} + \frac{\sqrt{11}}{32} Y_{7 4 c} + \frac{3\sqrt{143}}{32\sqrt{2}}
   Y_{7 6 c}   .
  \end{eqnarray}
    The equivalent deformation set for axially symmetric first and second
   minima follows from the expressions for spherical harmonics
   $Y^y_{\lambda 0}$:
  \begin{eqnarray}
   Y^y_{4 0} & = & \frac{3}{8} Y_{4 0}+ \frac{\sqrt{5}}{4} Y_{4 2 c} +
     \frac{\sqrt{35}}{8} Y_{4 4 c} ,  \\  \nonumber
   Y^y_{6 0} & = & -\frac{5}{16} Y_{6 0} - \frac{\sqrt{105}}{16\sqrt{2}}
     Y_{6 2 c} - \frac{3\sqrt{7}}{16} Y_{6 4 c} - \frac{\sqrt{231}}{16\sqrt{2}}
     Y_{6 6 c}  ,  \\  \nonumber
   Y^y_{8 0} & = & \frac{35}{128} Y_{8 0} + \frac{3\sqrt{35}}{32\sqrt{2}}
      Y_{8 2 c} + \frac{3\sqrt{77}}{64} Y_{8 4 c} + \frac{429}{32\sqrt{2}}
     Y_{8 6 c} + \frac{3\sqrt{715}}{128} Y_{8 8 c}  .
  \end{eqnarray}

\section{Dipole moments within the microscopic-macroscopic method}

  The macroscopic part of the expectation value of the electric diple moment
  is calculated as a sum of the redistribution and the neutron skin effects
  \cite{DMS}
  \begin{equation}
   {\bf D} = {\bf D}_{red} + {\bf D}_{skin}  ,
  \end{equation}
   where for $Z$ protons and $N$ neutrons, $A=Z+N$, $I=(N-Z)/A$, one obtains
  from the Droplet Model
  \begin{equation}
  {\bf D}_{red}= \frac{AZe^2}{8}\left(\frac{1}{J}+\frac{6LI}{KJ}\right)
  \left(<v>_V <{\bf \xi}>_V-<v{\bf \xi}>_V\right),
  \end{equation}
  \begin{eqnarray}
  {\bf D}_{skin} &  = & \frac{2NZ}{A}(I-{\bar \delta})R_0
  \left( <{\bf \xi}>_V-<{\bf \xi}>_S\right)  \\  \nonumber
  & + & \frac{9}{32}\frac{ZA^{2/3}e^2}{Q}B_S\left(<v>_S<{\bf \xi}>_S-
 <v{\bf \xi}>_S \right)   . \\ \nonumber
  \end{eqnarray}
  The Droplet Model parameters are: the volume symmetry-energy coefficient $J$,
 the nuclear incompressibility $K$, the effective neutron skin stiffness $Q$,
 the density symmetry coefficient $L$, the nuclear radius $R_0=r_0 A^{1/3}$
 and the equilibrium value of the average relative neutron excess,
 \begin{equation}
  {\bar \delta}=\frac{I+\frac{9 e^2}{80r_0Q}ZA^{-2/3}}
 {1+\frac{9J}{4Q}A^{-1/3}} .
  \end{equation}
 Nuclear shape enters through the constant $B_S$, being the ratio of the
 area of a deformed surface to the surface area of the sphere of the same
 volume, and the averages, over both the nuclear volume and surface,
 of the scaled
 radius vector ${\bf \xi}={\bf r}/R_0$ and the Coulomb potential $v$ in units
 of $Ze/R_0$, with $<f>_V$ meaning $\int_V f/V$, and $<f>_S$ meaning
 $\int_S f/S$. The calculations were performed with the following
  values of the parameters: $r_0=1.16$ fm, $J=$32.5 MeV, $K=$240 MeV,
 $Q=$50 MeV and $L=$100 MeV, as in \cite{dip}.

 The microscopic part of the dipole moment is calculated as $D^{micr}=
  \zeta (<{\hat D}> - {\tilde D})$ \cite{Lean,ButNaz},
 where the first part is the
  expectation value of the dipole operator on the deformed state and
  ${\tilde D}$ is the analogous expression in which the actual pairing
 occupation numbers $2 v_i^2$ were replaced by the quantities smoothed
 according to the Strutinsky prescription. The factor $\zeta$ takes care of the
 reduction of the effective charge due to the particle-vibration coupling to
 the $E1$ giant resonance. We have used $\zeta=0.33$ as it was done
 in previous calculations \cite{ButNaz,dip}.


\begin{thebibliography}{99}

\bibitem{Thirolf}
  P.~G.~Thirolf and D.~Habs, Prog. Part. Nucl. Phys. 49, 325 (2002);
  P.~Thirolf, D.Dc. Thesis, Ludwig-Maximilians-Universitat Munchen
  (2003).

\bibitem{Pol}
 S.~M.~Polikhanov et al., Sov. Phys. JETP 15, 1016 (1962).

\bibitem{FD66}
 G.~N.~Flerov and V.~A.~Druin, Preprint R-2539, JINR, Dubna
 (1966).

\bibitem{St67}
 V.~M.~Strutinsky, Nucl. Phys. A 95, 420 (1967).

\bibitem{Specht}
 H.~J.~Specht, J.~Weber, E.~Konecny and D.~Heunemann,
  Phys. Lett. B 41, 43 (1972).

\bibitem{Rev}
 V.~Metag, D.~Habs and H.~J.~Speth, Phys. Rep. 65, 1 (1980).

\bibitem{Backe80}
 H.~Backe et al., Phys. Rev. Lett. 80, 920 (1980).

\bibitem{Makar}
 V.~E.~Makarenko, INDC(CCP)-394 (1995).

\bibitem{NerV1}
  K.~Neergard and P.~Vogel, Nucl. Phys. A 145, 33 (1970),
    Nucl. Phys. A 149, 217 (1970).

\bibitem{RPAJ}
 T.~Nakatsukasa, K.~Matsuyanagi, K.~Mizutori and Y.~R.~Shimizu, Phys. Rev C 53,
  2213 (1996).

\bibitem{RPAR}
 V.G. Soloviev, A. Sushkov, and N. Yu. Shirikova, Z. Phys. A 358, 117 (1997).


\bibitem{WS}
  S.~\'Cwiok, J.~Dudek, W.~Nazarewicz, J.~Skalski and T.~Werner, Comput. Phys.
 Commun. 46, 379 (1987).

\bibitem{KN}
  H.~J.~Krappe, J.~R.~Nix and A.~J.~Sierk, Phys. Rev. C 20, 992
  (1979).

\bibitem{WSpar}
 M. Kowal, P. Jachimowicz, A. Sobiczewski, {\it Phys. Rev. C} {\bf 82}, 014303 (2010).

 \bibitem{Stefan}
 S.~\'Cwiok, W.~Nazarewicz, J~.X~.~Saladin, W.~P\l {}\'ociennik and A.~Johnson,
 Phys. Lett. B 322, 304 (1994).

\bibitem{Delar}
 J.-P.~Delaroche, M.~Girod, H.~Goutte and J.~Libert, Nucl. Phys. A 771, 103
 (2006).

\bibitem{collPom}
 L.~M.~Robledo, J.~L.~Egido, B.~Nerlo-Pomorska and K.~Pomorski,
  Phys. Lett. B 201, 409 (1988).

 \bibitem{BM}
 A.~Bohr, B.~R.~Mottelson, {\it Nuclear Structure} Vol. 2 (Benjamin,New York,1975)

\bibitem{Lean}
  G.~A.~Leander, W.~Nazarewicz, G.~F.~Bertsch and J.~Dudek, Nucl. Phys. A 453,
 58 (1986).

\bibitem{ButNaz}
  P.~A.~Butler and W.~Nazarewicz, Nucl. Phys. A 533, 249 (1991).

\bibitem{DMS}
 C.~O.~Dorso, W.~D.~Myers and W.~J.~Swiatecki, Nucl. Phys. A 451, 189
 (1986).

\bibitem{dip}
 J.~Skalski, Phys. Rev. C 49, 2011 (1994).

\bibitem{Schmorak}
M. Schmorak et al., {\it Nucl. Phys. A}  {\bf 178}, 410 (1972).

\bibitem{Firestone}
 ,,TABLE OF ISOTOPES'' edited by R. B. Firestone and V. S. Shirley,
 ISBN 0471-14918-7 Eight Edition, Vol 2, (1996).

 \bibitem{sletten}
 H.~C.~Britt, S.~C.~Burnett, B.~H.~Erkkila, J.~E.~Lynn
 and W.~E.~Stein, {\it Phys. Rev. C} {\bf 4}, 1444 (1971); G.~Sletten,
 V.~Metag and E.~Liukkonen, {\it Physics Letters B} {\bf 60}, 2 (1976).

\bibitem{240PuIIa}
 D.~Pansegrau et al, Phys. Lett. B 484, 1 (2000).

\bibitem{Gassmann}
D. Gassmann et al., {Physics Letters} {\bf B} 497, 3-4, 11 (2001).

\bibitem{230U}
 B.~Ackermann et al., Nucl. Phys. A 559, 61 (1993).

\bibitem{Zeyen}
P.~Zeyen et al, Z. Phys. A 328, 399 (1987).

\bibitem{Browne}
 E.~Browne and J.~K.~Tuli, Nuclear Data Sheets 107, 2649 (2006).

\bibitem{238U}
 K.~Mc Gowan and W.~T.~Milner, Nucl. Phys. A 571, 569 (1994);
  D.~Ward et al., Nucl. Phys. A 600, 88 (1996).

\bibitem{240Pu}
 I.~Wiedenh\"over et al., Phys. Rev. Lett. 83, 2143 (1999).

\bibitem{230Th}
Ch.~Lauterbach et al., Phys. Lett. B 130, 187 (1984).







%















\end{thebibliography}
\end{document}